\documentclass[
 reprint,
 amsmath, amssymb,
 prl,
 aps,
 superscriptaddress
]{revtex4-2}

\usepackage{graphicx}
\usepackage{subfigure}
\usepackage{hyperref}
\usepackage{bm}
\usepackage{float}
\usepackage{color}
\usepackage{amsmath}
\usepackage{soul}
\setstcolor{black}
\setulcolor{blue}

\hypersetup{colorlinks=true, citecolor=blue, urlcolor=blue, linkcolor=blue}
\begin{document}

\preprint{APS/123-QED}

\title{Self-Organized Time Crystal in Driven-Dissipative Quantum System}

\author{Ya-Xin Xiang}
\author{Qun-Li Lei}
      \affiliation{National Laboratory of Solid State Microstructures and School of Physics,
Collaborative Innovation Center of Advanced Microstructures, Nanjing University, Nanjing 210093, China}

\author{Zhengyang Bai}
    \email{zhybai@lps.ecnu.edu.cn}
    \affiliation{State Key Laboratory of Precision Spectroscopy, East China Normal University, Shanghai 200062, China}
  
\author{Yu-Qiang Ma}
     \email{myqiang@nju.edu.cn}
     \affiliation{National Laboratory of Solid State Microstructures and School of Physics,
Collaborative Innovation Center of Advanced Microstructures, Nanjing University, Nanjing 210093, China}

\begin{abstract}
Continuous time crystals (CTCs) are characterized by sustained oscillations that break the time translation symmetry. Since the ruling out of equilibrium CTCs by no-go theorems, the emergence of such dynamical phases has been observed in various driven-dissipative quantum platforms. The current understanding of CTCs is mainly based on mean-field (MF) theories, which fail to address the problem of whether the continuous time-translation symmetry can be broken in noisy, spatially extended systems absent in all-to-all couplings. Here, we propose a  CTC realized in a quantum contact model through self-organized bistability (SOB). The  CTCs stem from the interplay between collective dissipation induced by the first-order absorbing phase transitions (APTs) and slow constant driving provided by an incoherent pump. The stability of such oscillatory phases in finite dimensions under the action of intrinsic quantum fluctuations is scrutinized by the functional renormalization group method and numerical simulations. Occurring at the edge of quantum synchronization, the CTC phase exhibits an inherent period and amplitude with a coherence time linearly diverging with system size, thus also constituting a boundary time crystal (BTC). Our results serve as a solid route towards self-protected CTCs in strongly interacting open systems.
\end{abstract}

\maketitle

\emph{Introduction.}—Time crystals are self-organized spatiotemporal structures, first envisaged by Wilczek \cite{wilczek2012quantum,shapera2012classical}, that spontaneously break the time-translation symmetry imposed by underlying Hamiltonians. Since the advent of the no-go theorems stating that it is impossible to observe a spontaneously oscillating ground state (in thermal equilibrium)\cite{bruno2013impossibility,watanabe2015absence}, there have been several efforts concentrating on time crystals in closed Floquet systems \cite{khemani2016phase,Yao2017Discrete,choi2017obs, smits2018obs, Gambetta_Discrete_2019, bluvstein_controlling_2021}. Alternatively, coupling to an environment leads to the dissipative version of  time crystals that break the discrete/continuous time-translation symmetry of the dynamical generators  \cite{kessler2021obs,sarkar2022signatures,piazza2015self,chan2015limit,iemini2018boundary,Buca_Non_2019, Nie_Mode_2023,chen2023realization, Measurement_Krishna_2023}.

By building up a limit cycle (LC), the rise of synchronization in diverse physical platforms, such as optomechanical oscillators~\cite{Sheng_Self2020}, Rydberg gases~\cite{Ding2023Ergodicity,wu2023observation,wadenpfuhl2023emergence} and hybrid atom-cavity systems~\cite{kessler2019emergent,kongkhambut2022obs}, has been observed and related to the formation of CTCs. In open systems, the dissipation often associates with the quantum Langevin noise, it is probable that the fluctuations would affect the robustness of CTCs thereby destroying the crystalline order. Notwithstanding the rapid advances in experimental studies, to what extend the CTCs remain intact under the action of intrinsic noise is an open question worthy of theoretical endeavors. 

Analogous to the famous notion of self-organized criticality (SOC), which is related to self-organization to the critical point of a continuous APT~\cite{tang1988mean, sornette1992critical, paczuski1994field, grinstein1995, gil1996landau, dicman1998self, dickman2000paths, buendia2020feedback}, the mechanism for SOB consists in a separation of the time scale of the dynamics of the order parameter from that of the corresponding control parameter. It triggers a LC phase of the hysteresis loop of a first-order APT~\cite{disanto2016sob, buendia2020feedback}. In light of the common features shared by LCs and CTCs, a new class of CTCs induced by SOB can be envisioned.

\begin{figure}[t]
	\centering
	\includegraphics[width = 8.6 cm, keepaspectratio]{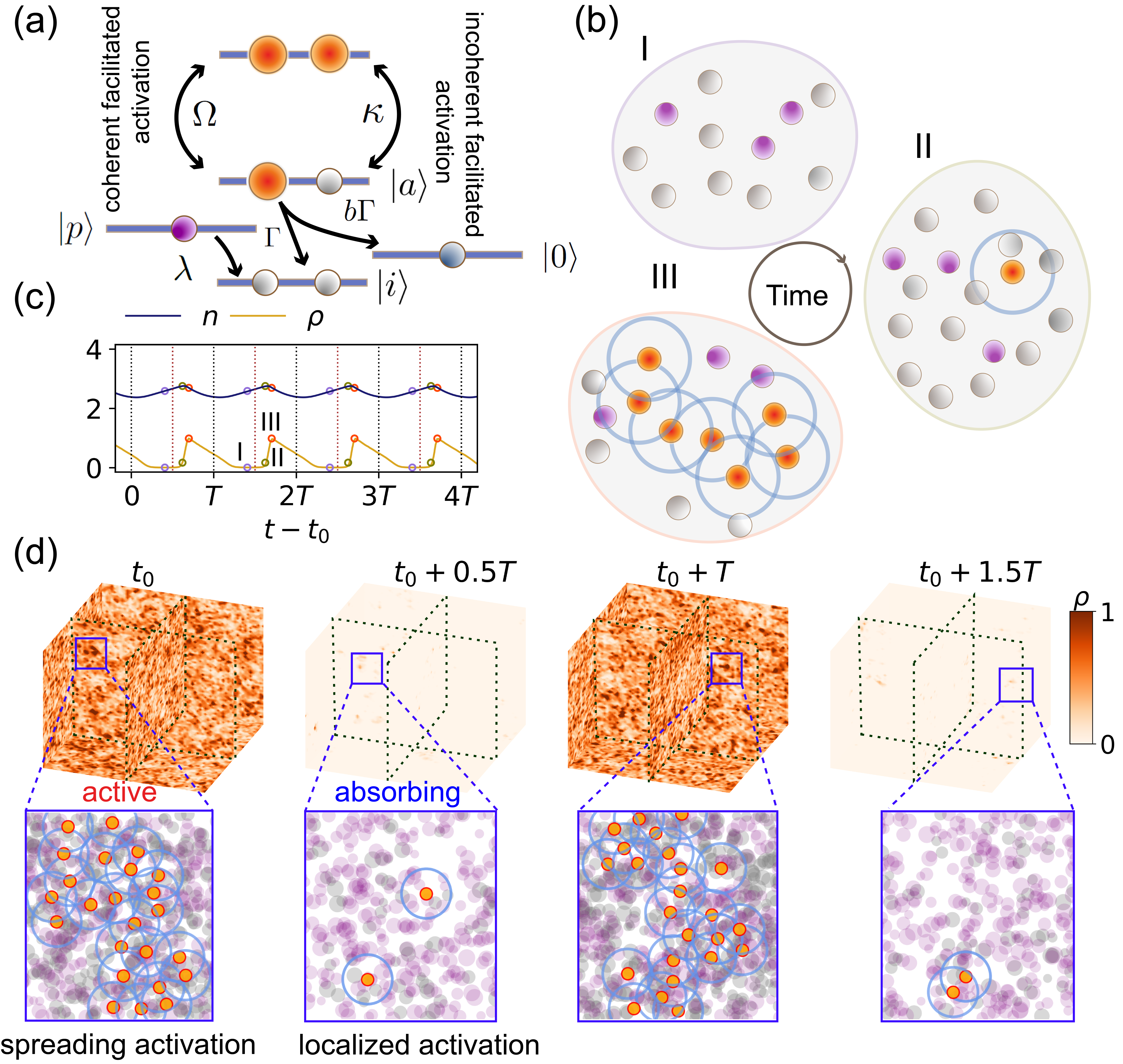}
	\caption{(a) The effective four-level scheme. The quantum emitter in inactive state $|i\rangle$ (gray sphere) in proximity to emitters in active state $|a\rangle$ (red sphere) can become active via (in)coherent facilitated activation, and active emitters can spontaneously decay into either inactive or removed states $|0\rangle$ (blue sphere). Emitters in $|p\rangle$ state (purple sphere) are incoherently pumped to  inactive state $|i\rangle$. (b) Sketch of  CTC, where the contact activation occurs on the facilitation shell (large blue hollow sphere) of active emitters. The system typically consists of subcritical (I) and supercritical (II) states of low active densities, and supercritical highly active states (III). (c) Dynamical oscillations of the average total $n$ and active $\rho$ densities, corresponding to three states in panel (b). (d) Snapshots of the active density field from simulations are sketched.  %Parameters are $\kappa=0,\Omega=0.5, \lambda=3.2\times10^{-3}, L = 64$.
	}
	\label{fig:model}
\end{figure}

In this work, we theoretically investigate the formation and stability of CTCs beyond MF approximation. Concretely, we consider a dissipative variant of the contact model characterized by the quantum and classical contact interactions between quantum emitters~\cite{marcuzzi2016absorbing}. In the classical regime, the system undergoes continuous APTs. However, the transitions become discontinuous in the quantum regime. Upon addition of slow loading mechanism, a non-stationary phase arises from SOB, where the number of quantum emitters changes periodically [numerical results for three-dimensional systems sketched in Fig.~\ref{fig:model}(c)-(d)]. Meanwhile, the system undergoes repeated phase transitions and self-organizes to a CTC phase. Avalanches of activation, like fires spreading in a forest, trigger collective jumps from the absorbing to the active states, and terminate upon the exhaustion of emitters (trees), which in turn bring the system back into the absorbing state of slow recovery, waiting for the next jump. The CTC here is analogous to the breathing mode of ``forest-fires'' model~\cite{Ding_Phase_2020}. Through theoretical analysis and numerical simulations, we find that the CTCs are unstable in low-dimensional systems, due to the reduced effective barrier separating the active from the absorbing states. Besides, our CTC suffers from phase noises caused by the kinetic trapping in absorbing state. Consequently, the coherence time diverges linearly with system size, whereas the degree of quantum synchronization decreases in larger systems. 
%Possible experimental implementations are briefly discussed in the end.

\emph{Model.}—Our model can be represented as an effective four-level system [Fig.~\ref{fig:model}(a)], where quantum emitters in  active state $|a\rangle$ can spontaneously decay into  the inactive state $|i\rangle$ (with rate $\Gamma$), and the inactive ones can be activated only in the vicinity of active ones both incoherently and coherently (with rates $\kappa$ and $\Omega$). Additionally, loss of emitters due to the decaying of active emitters into the removed states $|0\rangle$ (with rate $b\Gamma$), and an incoherent coupling $|p\rangle\to|i\rangle$ that mimics injecting inactive emitters (with rate $\lambda$) are included. In free space, we can restrict the contact processes to pairs of emitters that are separated by the facilitation radius $R_{\rm fac}$, and refer to the effective nearest neighbors of an active emitter as emitters at the border of its facilitation sphere %[large blue hollow spheres in Fig. \ref{fig:model}(b)]
\cite{helmrich2020signatures}. 

Under the Markovian noise, the effective dynamics of this system permits a microscopic description for the density operator $\hat\rho$ via a Lindblad master equation $\partial_t \hat\rho = -i[\hat H,\hat\rho] + \sum_{\alpha} \mathcal{L}_{\alpha}\hat\rho$.
The coherent activation is described by the effective Hamiltonian ($\hbar\equiv1$ henceforth)
\begin{equation}
	\hat H = \Omega \sum_l {\hat C}_l {\hat\sigma}_l^x,
\end{equation}
where ${\hat C}_l = \sum_{k\in\partial l}{{\hat\sigma}_k^{aa}}, {\hat\sigma}_l^{\alpha\beta}\equiv|\alpha_l\rangle\langle \beta_l|$ ($\alpha, \beta = a, i, p, 0$), $l,k$ are indices for each emitter, and $\sum_{k\in\partial l}$ denotes a summation of the effective nearest neighbors of the $l$-th emitter. The operator ${\hat\sigma}_l^x=\hat\sigma^{-}_l+\hat\sigma^{+}_l$ flips the quantum state with the ladder operators $\hat\sigma^{+}_l \equiv \hat\sigma_l^{ai}$ and $\hat\sigma^{-}_l \equiv \hat\sigma_l^{ia}$.

The dissipative dynamics is described by Lindblad terms $\mathcal{L}_{\alpha}\hat\rho = \sum_{l} \left[{\hat L}_{\alpha,l}\hat\rho {\hat L}_{\alpha,l}^\dagger - \frac{1}{2}\left\{{\hat L}_{\alpha, l}^{\dagger}{\hat L}_{\alpha, l},\hat\rho \right\}\right]$. The spontaneous inactivation of the active states is described by ${\hat L}_{d,l} = \sqrt{\Gamma}\hat\sigma^{-}_l$, and ${\hat L}_{p,l} = \sqrt{\gamma_{de}}{\hat\sigma}_l^{aa}$ represents dephasing of quantum coherence with rate $\gamma_{de}$. 
Meanwhile, the loss and reloading of inactive emitters are accounted for by ${\hat L}_{e,l} = \sqrt{b\Gamma}\hat\sigma_l^{0a}$ and ${\hat L}_{a,l}=\sqrt{\lambda}\hat\sigma_l^{ip}$, respectively. The incoherent contact processes are also included in Lindbladian, where the respective jump operators for activation and inactivation of emitters are given by ${\hat L}_{b,l} = \sqrt{\kappa}{\hat C}_l{\hat\sigma}_l^{+}$ and ${\hat L}_{c,l} = \sqrt{\kappa}{\hat C}_l{\hat\sigma}_l^{-}$. 

% \bblue{The loss, reloading, incoherent activation and inactivation of inactive emitters are accounted for by ${\hat L}_{e,l} = \sqrt{b\Gamma}\hat\sigma_l^{0a}$, ${\hat L}_{a,l}=\sqrt{\lambda}\hat\sigma_l^{ip}$, ${\hat L}_{b,l} = \sqrt{\kappa}{\hat C}_l{\hat\sigma}_l^{+}$ and ${\hat L}_{c,l} = \sqrt{\kappa}{\hat C}_l{\hat\sigma}_l^{-}$, respectively.}

The Heisenberg-Langevin equations of motion for the operators $\hat\sigma^{x/aa}_l$, $\hat\sigma^{y}_l=i{\hat\sigma}_l^{-}-i{\hat\sigma}_l^{+}$, and $\hat n_l=\hat\sigma_l^{aa}+\hat\sigma_l^{ii}$ 
%according to the master equation 
read
\begin{subequations}
	\begin{align}
		\partial_t \hat\sigma_l^{aa}  = &  -\Gamma\hat\sigma_l^{aa} + \Omega{\hat C}_l\hat\sigma_l^{y} + \kappa\hat C_l\left(\hat n_l - 2\hat\sigma_l^{aa}\right) + \hat\xi_l^{aa} \label{eq:EOM_qm_rr}\\
		\partial_t \hat\sigma_l^{x} = &-\frac{\kappa\hat N_l + \gamma}{2} \hat\sigma^x_l - \kappa\hat C_l\hat\sigma_l^{x} - \Omega{\hat P}_l\hat\sigma_l^{y} + \hat\xi_l^{x} \label{eq:EOM_qm_x}\\
		\partial_t \hat\sigma_l^{y} =&  -\frac{\kappa{\hat N}_l + \gamma}{2} \hat\sigma^y_l - \kappa{\hat C_l}\hat\sigma_l^{y} + \Omega{\hat P}_l\hat\sigma_l^{x} \label{eq:EOM_qm_y}\\
		&+ 2\Omega\hat C_l\left(\hat n_l - 2\hat\sigma_l^{aa}\right) + \hat\xi_l^{y}\notag\\
		\partial_t \hat n_l =& -b\Gamma\hat\sigma_l^{aa} + \lambda\hat\sigma_l^{pp}+ \hat\xi^{n}_l \label{eq:EOM_qm_n}
	\end{align}
\end{subequations}
where ${\hat N}_l =  \sum_{k \in\partial l} {{\hat n}_k} $, ${\hat P}_l = \sum_{k\in\partial l} {{\hat\sigma}_k^{x}}$, and $\gamma = \Gamma + \gamma_{de}$. The Langevin noise operators $\hat\xi_l^{x/y/aa/n}$ appear because the dissipation is attributed to the coupling between the system and a large reservoir \cite{pan2020non}, and can by fixed via solving the Heisenberg equations under the reservoir Hamiltonians in Born-Markov approximation [See  Supplemental Material ({\bf SM}) for details]. Thereafter, we set the time unit to $\Gamma^{-1} = 1$.

% This method allows us to obtain the correct form of the noise in consistency with the deterministic equations.

In the following, we consider the continuum limit and after coarse-graining transform the expectation values of the operators into classical fields. More specifically, the fields of active and total densities are defined as $\rho\left(\boldsymbol{r},t\right) \equiv\left\langle \text{Tr}\left\{\hat\sigma^{aa}_l\hat\rho \right\}\right\rangle_{\boldsymbol{r}}$ and $n\left(\boldsymbol{r},t\right) \equiv\left\langle\text{Tr}\left\{\hat n_l\hat\rho \right\}\right\rangle_{\boldsymbol{r}}$, respectively, where $\langle...\rangle_{\boldsymbol{r}}$ denotes an average over the facilitation sphere centered at $\boldsymbol{r}$, 
and %the coherence fields 
$\sigma^{x/y}$ are defined similarly. Omitting the operator moments generated by the two-body interactions and perturbatively eliminating $\sigma^{x/y}$ via the Janssen-De Dominicis-Martin-Siggia-Rose procedure \cite{Janssen1976, Martin1973} (see {\bf SM}) lead to the following action for the active density field
\begin{equation}
\label{eq:S_rho}
	S[\rho, \tilde\rho] = \int {\tilde\rho\left[\left(\partial_t - D_\rho\nabla^2 + u_2\right)\rho  + u_3 \rho^2 +u_4\rho^3- \frac{\mu}{2}\tilde\rho\right]}
\end{equation}
where $\tilde\rho$ is the Martin-Siggia-Rose auxiliary field related to the dynamic responses of $\rho$ to perturbations, and $u_2 = 1 - n\kappa - 256 n^2\Omega^4/\left(n\kappa + \gamma\right)^7, u_3 = 2\left[\kappa - 2n\Omega^2/(n\kappa + \gamma)\right], u_4 = 8\Omega^2/(n\kappa + \gamma)$, $\mu = \left(1 + n\kappa\right)\rho + 4n\Omega^2 \rho^2/\left(n\kappa + \gamma\right)^2$ are the coupling constants. The diffusion constant $D_\rho = D_\text{T} + n\kappa R_\text{fac}^2 / 2$, where $D_\text{T}$ is the thermal diffusivity. 
%\textcolor{red}{It's worth noting that fluctuations of the coherence fields destabilize the absorbing phase on top of the classical contact processes at high total densities. As will be shown later, in the quantum regime ($\Omega\gg\kappa\approx0$), the noise is not only responsible for the bistability but causing the transitions between them as well, thus enabling CTC to occur.}
% we shall use ``bistability'' and the first-order APTs interchangeably throughout the letter.

\begin{figure}[t]
	\centering
	\includegraphics[width = 8.6 cm, keepaspectratio]{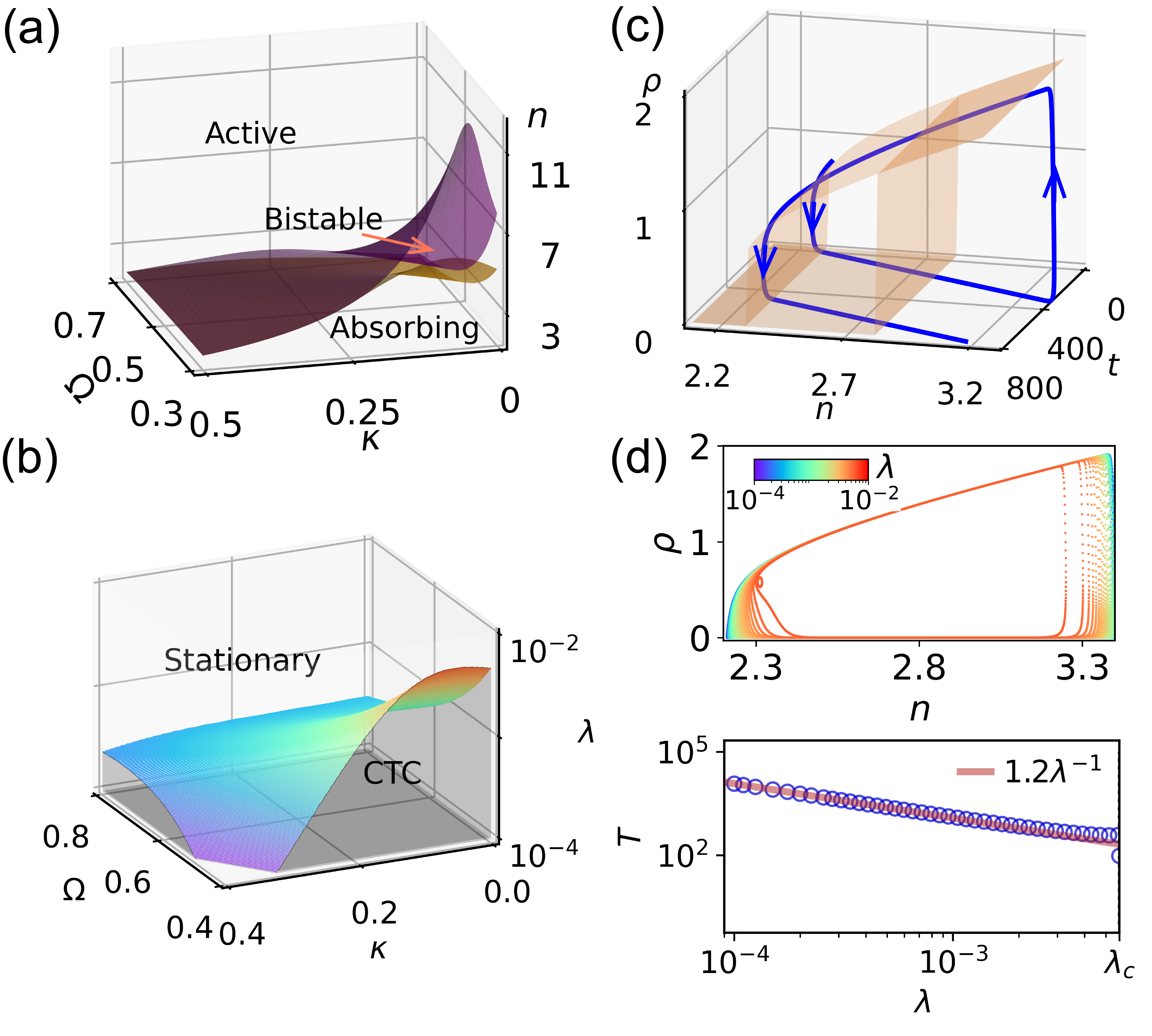}
	\caption{Phase diagrams. (a) The two surfaces are the phase boundaries. Discontinuous APTs ensue when the two surfaces are separated from each other. (b) The surface represents the critical rate $\lambda_c$ (color-coded) that separates a  stationary phase from a CTC phase, corresponding to stable and unstable fixed points. (c) The CTC phase that consists of self-organized jumps (blue arrows) between the active and the absorbing phases (orange surfaces) for $\lambda=3.2\times10^{-3}$. (d) Phase-space trajectories (upper) and periods (lower) as a function of $\lambda$. %Parameters for (b) and (d) are $\Omega=0.5, \kappa = 0$.
	}
	\label{fig:mf}
\end{figure}

%\textcolor{red}{We first construct the phase diagrams.} 
With the total density $n$ conserved ($b,\lambda=0$), the static phases are determined by the solutions to the saddle-point equations following variation of action \eqref{eq:S_rho} with $\tilde\rho,\tilde n=0,D_\text{T}\to\infty$. The corresponding phase boundaries are shown in Fig.~\ref{fig:mf}(a). In the quantum regime ($\Omega\gg\kappa$), the bistable region within the two boundaries indicates that the systems undergo discontinuous APTs when the total density $n$ exceeds a critical value (see {\bf SM} for  relevant phase diagrams), which is an element of SOB-induced CTCs to be discussed later. Approaching the classical regime ($\Omega=0$), the bistable region shrinks and finally vanishes when the transition becomes continuous, which has been related to SOC in driven-dissipative Rydberg gases \cite{klocke2019controlling,helmrich2020signatures, Ding_Phase_2020}. 

\emph{SOB-induced CTCs.}—Having identified the regime for first-order APTs, sustained oscillations can arise from the interplay between loss and reloading of emitters ($b,\lambda\neq0$). Our proposal for SOB-induced CTCs is encoded in the Langevin equations for the density fields $\rho,n$ as follows
\begin{subequations}\label{Dyn}
	\begin{align}
		&\partial_t \rho = D_\rho\nabla^2\rho +\tau n - u_2\rho- u_3\rho^2 - u_4\rho^3 + \eta\label{eq:drhodt}\\
		&\partial_t n = D_T\nabla^2 n - b\rho + \lambda + \xi^n\label{eq:dndt}
	\end{align}
\end{subequations}
where $\eta, \xi^n$ are Markovian white noises with vanishing mean and respective variance $\mu+\tau n$ and $b\rho$. To prevent the system from trapping in absorbing states, a small driving $\tau n$ is added. %Previous studies of CTCs in atomic gases focused on the $D_\text{T}\to\infty$ limit, where spatial inhomogeneities are suppressed, and saddle-point equations are accurate. In our case, we work with finite diffusivity. 
Throughout the paper, we fix $\tau=10^{-7}$,  $b=0.01$, $D_\text{T}=1$, $n_p=1$, $\kappa=0$, $\Omega=0.5$ and $\gamma=2$, unless otherwise stated. 

% ~\cite{wadenpfuhl2023emergence, wu2023observation} Our analysis is not exactly MF, because we haved used the dynamic path integral formalism to average out all fast degrees of freedom instead of neglecting them.

In the presence of first-order APTs [shaded region in Fig. \ref{fig:mf}(b)], one can identify a critical loading rate $\lambda_c$ %[surface plotted in Fig. \ref{fig:mf}(b)] 
through a linear stability analysis of the fixed point possessed by Eqs. \eqref{eq:drhodt} and \eqref{eq:dndt}. The correlation functions diverge at a finite frequency, signaling the onset of broken time translation symmetry \cite{lee2011anti,scarlatella2019emergent,nie2023mode} [see {\bf SM}]. The CTCs feature alternating jumps between the states with low and high active densities [Fig.~\ref{fig:mf}(c)] with a period  $\propto\lambda^{-1}$ [Fig.~\ref{fig:mf}(d)]. The remaining  is devoted to a systematic study of the CTC in finte systems.
%\st{below which LC phases that break time translation symmetry arise from the Hopf bifurcation. The existence of such oscillatory phases requires first-order APTs [shaded region in Fig. \ref{fig:mf}(c)], because for systems with continuous APTs, the fixed points are always stable, and STA states are expected. Results of numerical integration of the MF equations are displayed in Fig. \ref{fig:mf}(d),where the SOB-induced LCs alternate between two phases. } 

% bifurcation theory \cite{cross1993pattern,lefever1971chemical,lee2011anti} and roton instabilities are the same here.

\begin{figure}[b]
	\centering
	\includegraphics[width = 8.6 cm, keepaspectratio]{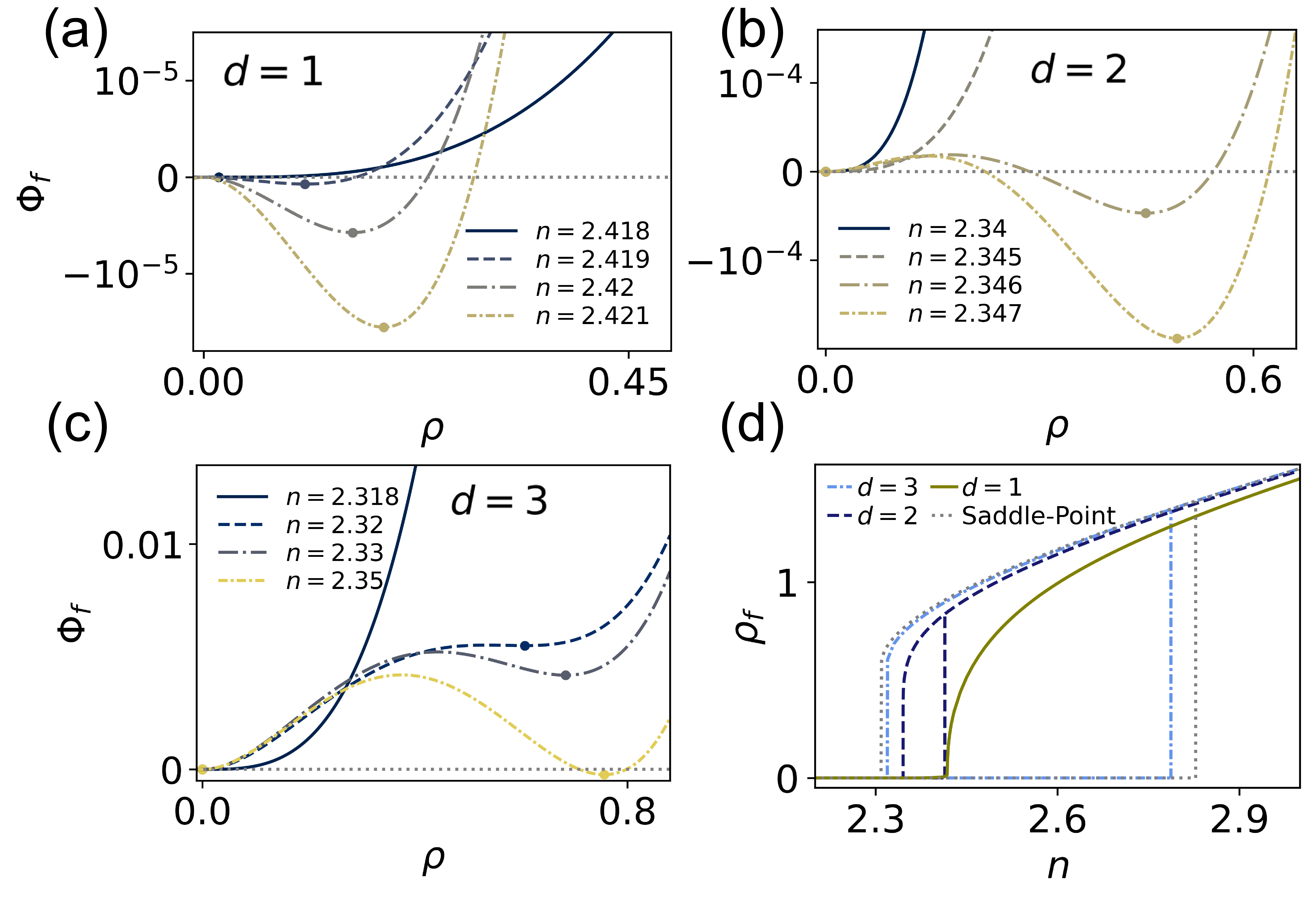}
	\caption{The effective potential $\Phi_f$ as a function of the active density field $\rho$ for different $n$ in (a) $d=1$, (b) $d=2$, (c) $d=3$. The active/absorbing phases are determined by the local minima (dots). (d) The corresponding phase diagram in comparison with that obtained via saddle-point approximation.
% The result for $d=3$ approaches the MF line. 
%Parameters for (a)-(d) are $\kappa=0, \Omega=0.5$. 
}
	\label{fig:NPRG}
\end{figure}

%\textcolor{red}{(a) Time series of the total and active densities.  Upper row: $\lambda = 8\times 10^{-4}$ with $d=1, L=10^{4}$ (left) and $d = 3, L=64$ (right); lower row: $\lambda = 1.2\times10^{-3}$ with $d=2, L=256$ (left) and $d = 3, L=64$ (right). (b) three-dimensional systems with $L = 24$ (left) and $L = 64$ (right).}

\begin{figure*}[t]
	\centering
	\includegraphics[width = 17.2 cm, keepaspectratio]{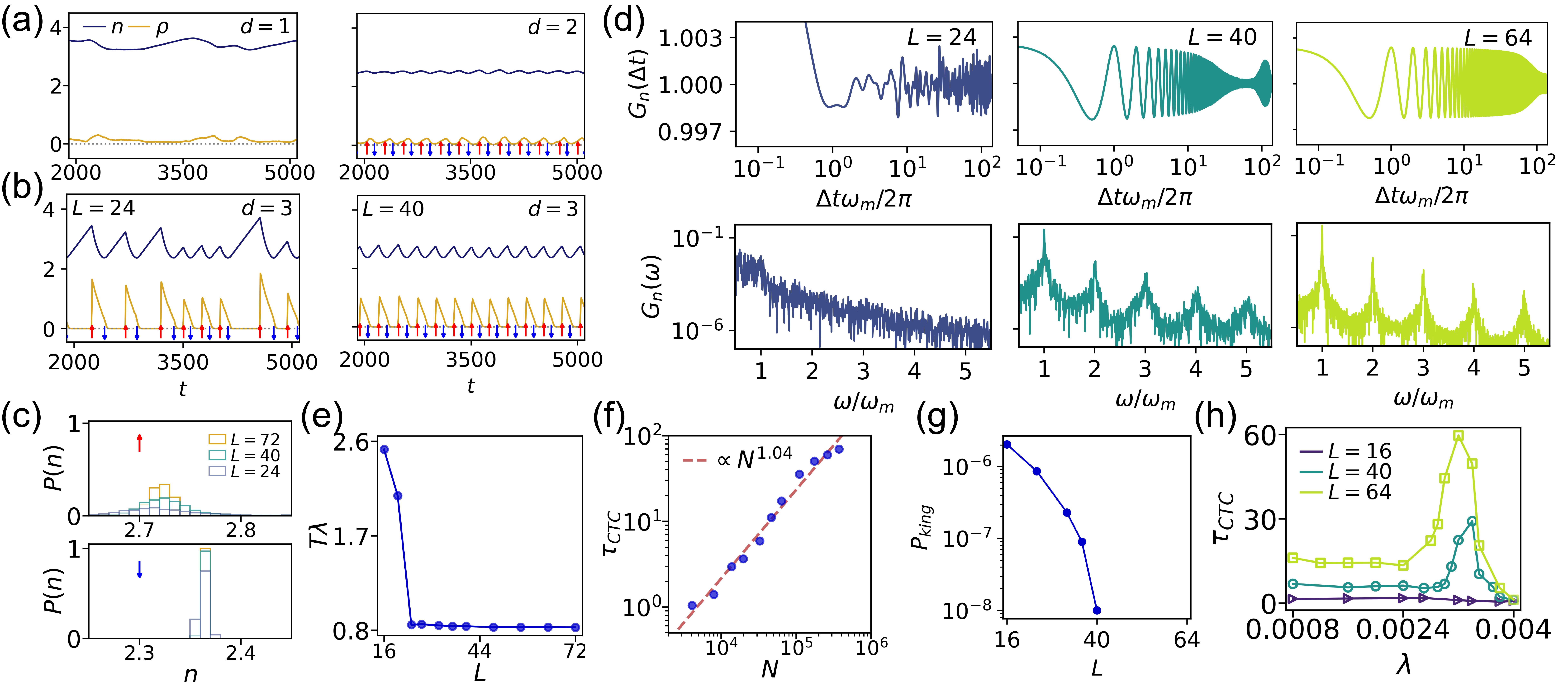}
	\caption{Time series of the densities with collective jumps marked by red (blue) arrows in (a) low dimensions for $\lambda=1.2\times10^{-3}$ with system size $L=10^{4}$ (left), $L=256$ (right) and (b) three-dimensional systems. (c) Probability distribution of the average total densities at which upward (upper) and downward (lower) jumps occur for varying $L$. (d) The autocorrelation functions $G_{n}(\Delta t)$ and their Fourier spectra $G_{n}(\omega)$, where the first peak $\omega_{m}$ dictates (e) the period $T=2\pi/\omega_m$. %(e) Rescaled period defined as the inverse of the location of the highest peak of the Fourier spectra. 
	(f) and (h) are the coherence time $\tau_\text{CTC}$ defined as the ratio of the period to its standard deviation, where $N=L^{3}$ is the system volume. (g) The occurrence probability of king avalanches. We select $d = 3$ [for (b)-(h)] and $\lambda=3.2\times10^{-3}$ [for (b)-(g)].}	
	%	(e) Rescaled period defined as the inverse of the location of the highest peak of the Fourier spectra. (f) The coherence time $\tau_\text{CTC}$ defined as the ratio of the period to its standard deviation versus the volume of the system $N=L^{3}$. (g) The occurrence probability of king avalanches for different $L$. (h) The coherence time as  functions of $L$  and $\lambda$. }
%\st{$\tau_\text{CTC}\equiv\pi/\Delta\omega$, with the half-width $\Delta \omega$ determined by Lorentz fit of the highest spectral peak.} 
	\label{fig:SIM}
\end{figure*}

\emph{Dimensionality dependence of CTC}—We first discuss how dimensionality affects the stability of the CTCs. In lower dimensions, the effective barrier between the two states can be reduced. The jumps might occur at a wider range of total densities, thus destroying the long-range time crystalline order. To acquire a quantitative description, we adopt a functional renormalization group (fRG) approach via the Wetterich equation \cite{wetterich1993exact, Canet2011General,Dupuis2021thenon} to obtain the flows of the phase structure with decreasing infrared cutoff for various values of $n$ in different dimensions (see {\bf SM}). The resulting effective potential $\Phi_f(\rho)$ and the corresponding phase diagram are shown in Fig. \ref{fig:NPRG}. In $d=1$, as $n$ increases, the position of the local minimum shifts continuously from the origin to a finite value, indicating a continuous transition [Fig. \ref{fig:NPRG}(a)]. In $d\geq2$, however, increasing $n$ induces the appearance of a second local minimum at the finite density, apart from the local minimum at the origin, with a barrier in between, indicating a first-order transition [Fig. \ref{fig:NPRG}(b) and (c)]. Besides, the barrier is higher in $d=3$ than $d=2$, suggesting a weaker first-order transition in lower dimensions. The phase diagram in accord is shown in Fig. \ref{fig:NPRG}(d), where we can infer that discontinuous transitions are expected for $d\geq2$. Compared with the MF results, the coexistence region becomes narrower for a lower dimensionality. The fRG results indicate that CTCs are possible in $d\geq2$. The three-dimensional CTC is protected by a higher barrier and thus more stable.
%, with the three-dimensional CTCs being protected by a higher barrier and thus more stable %between the two coexisting phases and thus more stable compared with the two-dimensional ones. 

We then numerically simulate Eqs. \eqref{eq:drhodt} and \eqref{eq:dndt} deploying the operator-splitting scheme \cite{pechenik1999interfacial, dornic2005integration}, and record the time-series of the average 
%total and active 
densities, from which we detect collective quantum jumps. 
%As shown in Fig. \ref{fig:SIM}(a), 
Deep in the CTC regime, oscillatory phases that entail a succession of periodic jumps are observed for $d=2$ [Fig. \ref{fig:SIM}(a), left panel]. Whereas compared with those in three-dimensional systems [Fig. \ref{fig:SIM}(b), right panel], the jumps in two-dimensional occur on a much lesser scale, in accord with the much weaker first-order APTs therein, as revealed by the fRG analysis. Given the enhanced stability of CTCs in higher dimensionalities,  in what follows, we restrict our discussion to $d=3$. Results for $d=1,2$ are given in {\bf SM}.

\emph{Finite-size effects.—} As evinced in Fig. \ref{fig:SIM}(b), more stable CTCs are expected in larger systems. Meanwhile, the distribution of average total density converges  as the system is enlarged [Fig. \ref{fig:SIM}(c)]. The two-time correlation functions $G_{x}(\Delta t)\equiv\left\langle x(t)\right\rangle_t^{-2} \left\langle x(t)x(t+\Delta t)\right\rangle_t$, for $x=n,\rho$ \cite{wu2023observation} manifest constant periodic oscillations for perfect time crystals\cite{wu2023observation} .
%and otherwise exhibit modulations at times due to decoherence.
Correspondingly, their Fourier spectra $G_x(\omega)$ peak at the integer multiples of their respective inherent frequencies $\omega_m$. We can infer from Fig. \ref{fig:SIM}(d), that in larger systems, the amplitude of $G_n(\Delta t)$ varies more slowly, and the Fourier spectra are more sharply peaked at $\omega/\omega_m=1,2,3…$, typical of periodic structures in time.
Besides, once the time crystalline order is built, the period $T\equiv 2\pi/\omega_m$ remains invariant with diverging $L$ [Fig. \ref{fig:SIM}(e)], and is thus inherent to CTCs. 
%\st{To quantify time crystalline order, we follow \cite{chen2023realization} to estimate the coherence time $\tau_\text{CTC}\equiv\pi/\Delta\omega$, where the half-width $\Delta\omega$ is determined by a Lorentz fitting of the Fourier spectra.} 
%\st{As we can infer from Fig. \ref{fig:SIM}(f), albeit finite, the coherence time increases with system size, suggesting the occurrence of persistent oscillations in the $L\to\infty$ limit. }
The existence of sustained oscillations with an intrinsic amplitude and frequency suggests that our CTCs are also a realization of BTCs \cite{iemini2018boundary}.

Since upward jumps always lead to correlations among distant emitters (avalanches), the enhanced time crystallinity cannot be explained through the method of system-size expansion \cite{kampen2007stochastic,boland2008how,chan2015limit,cabot2022quantum}. 
%Our simulations reveal the finite-size effect, which  %   \textcolor{red}{Rather, fluctuations are increasingly periodic and finite in enlarged systems (see {\bf SM}).} Because upward jumps trigger avalanches that lead to correlations among distant emitters.
The irregularity in the time series features a significant increase in the total density followed by an abrupt decrease in the active and total densities [see Fig. \ref{fig:SIM}(b), left panel]. For low loading rates, such events have been attributed to the system falling into the absorbing state, and the consequent overloading in turn brings about system-spanning activation avalanches \cite{grassberger2002critical, kinouchi2019stochastic}. 
%We argue that the same reasoning applies to the aperiodic oscillations in small systems therein. 
For a three-dimensional CTC with a volume of $N=L^{3}$, the accumulated phase shift $\Delta\theta$ per cycle is approximately the probability of kinetic trapping in the absorbing phase $\sim 1/\left(TN\tau\right)$, and CTC should remain coherent over the time scale (rescaled by $T$) $\tau_\text{CTC}\approx2\pi T/\Delta\theta$. We then estimates the coherence time from simulations by extracting the phase shifts via $\Delta\theta=2\pi \left\langle\Delta T\right\rangle/ T$, where $\left\langle \Delta T \right\rangle$ is the standard deviation of the period. Results are plotted in Fig \ref{fig:SIM}(f), where the coherence time increases nearly linearly with the volume with a prefactor of $1.5(4)\times 10^{-4}$. It's easy to reconcile the finite correlations and diverging coherence time in the $N\to\infty$ limit, because the absorbing state is devoid of fluctuations and has a lifetime inversely proportional to $N$. 
	%The sensitivity to metastable states is unique to CTCs, whereas normal crystals are equilibrium species, and therefore independent of any kinetic effect.

Also, the next upward jump following the kinetic trapping is likely to trigger huge avalanches. To test this idea, we count space-time activation avalanches by connecting sites with active densities larger than a threshold ($\tau$) as neighbors in the time-forward direction and grouping them into clusters. The occurrence probability of huge avalanches (king avalanches, defined as those  containing more than half the total number of sites) decreases as the system becomes larger [Fig. \ref{fig:SIM}(g)], in line with the longer coherence time [Fig. \ref{fig:SIM}(f)]. A comparison of $\tau_\text{CTC}$ among various loading rates and system sizes is displayed in Fig. \ref{fig:SIM}(h), the regime for CTCs indicated by the significantly increased coherence time lies between that for the aperiodic oscillations and the fluctuating uniform ones, and widens in larger systems. 

Frequent huge avalanches induce coherent changes in active and total densities among a great many sites and thus reflect the underlying quantum synchrony at its highest level. However, a lack of synchrony results in stationary states with small fluctuations that conserve time-translation invariance. Sustained periodic oscillations reside in between the above two scenarios, where discontinuous phase transitions spontaneously generate finite-range correlations, which are enough to trigger coherence among local sites and yet unable to support a global synchronization in infinite systems. In other words, SOB-induced CTCs arise at the edge of quantum synchronization.

\emph{Conclusion and discussion.}—In this work, we propose a mechanism to realize self-protected CTCs with diffusive couplings. Our analysis is not restricted to APTs, and can be generalized to other systems with bi-/multi-stability.  
Our CTCs can be interpreted as a BTC \cite{iemini2018boundary}, where the reservoir plays the role analogous to the bulk Hamiltonian.
% such that tracing out the corresponding degrees of freedom yields Lindbladian for the (boundary) degrees of freedom, where sustained oscillations with intrinsic amplitudes and periods emerge in the thermodynamic limit. 
%
%
The model here can be implemented with coherent laser-driven Rydberg atoms in anti-blockade regime, where the electronic ground (Rydberg) states can be mapped to the inactive (active) states~\cite{marcuzzi2016absorbing,klocke2019controlling,helmrich2020signatures}. The relative importance of the coherent and incoherent activation processes can be controlled by driven lasers.

Revealing how temporal organization arises from the SOB-induced bifurcation in quantum many-body systems, our study extends the dynamical phase diagram for both SOB and instability-related LCs \cite{epstein1996nonlinear, lefever1971chemical, buendia2020feedback}. Alongside the coherent-state path integral formalism for reaction-diffusion systems \cite{doi1976stochastic, peliti1976path, wiese2016coherent}, the  procedures facilitate studying real-world critical-like events \cite{malamud1998forest, dickman2000paths, yoshioka2003sandpile, beggs2003neuronal,  grassberger2002critical, bonachela2009self} through controllable platforms.

\begin{acknowledgments}
This work is supported by the National Natural Science Foundation of China under Grants No. 11974175, No. 12247102 and No. 12274131. We are grateful to the High Performance Computing Center (HPCC) of Nanjing University for performing the numerical calculations in this paper on its blade cluster system.
\end{acknowledgments}

\bibliographystyle{apsrev4-2}
%\bibliography{ref}
%apsrev4-2.bst 2019-01-14 (MD) hand-edited version of apsrev4-1.bst
%Control: key (0)
%Control: author (72) initials jnrlst
%Control: editor formatted (1) identically to author
%Control: production of article title (-1) disabled
%Control: page (0) single
%Control: year (1) truncated
%Control: production of eprint (0) enabled
%

\end{document}